\documentstyle[12pt]{article}
\begin{document}
%\draft

% below commands are for enumerating eqs according to sections
%\newcounter{eq}[section]
%\newcommand{\set}{\stepcounter{eq}
%\renewcommand{\theequation}{\mbox{\arabic{section}.\arabic{eq}}}}

\hsize=6.15in
\vsize=8.2in
\hoffset=-0.42in
\voffset=-0.3435in

\normalbaselineskip=24pt\normalbaselines

\begin{center}
{\large \bf Scaling of brain metabolism and blood flow in relation to 
capillary and neural scaling}
\end{center}

\vspace{0.15cm}

\begin{center}
{Jan Karbowski}
\end{center}

\vspace{0.05cm}

\begin{center}
{\it Institute of Biocybernetics and Biomedical Engineering, \\
Polish Academy of Sciences,
02-109 Warsaw, Poland }
\end{center}

%\date{\today}

\vspace{0.1cm}

%\widetext
\begin{abstract}
Brain is one of the most energy demanding organs in mammals, and its total 
metabolic rate scales with brain volume raised to a power of around 5/6. 
This value is significantly higher than the more common exponent 3/4 
relating whole body resting metabolism with body mass and several other 
physiological variables in animals and plants. This article investigates 
the reasons for brain allometric distinction on a level of its microvessels.
Based on collected empirical data it is found that regional cerebral blood 
flow CBF across gray matter scales with cortical volume $V$ as 
$\mbox{CBF} \sim V^{-1/6}$, brain capillary diameter increases as $V^{1/12}$, 
and density of capillary length decreases as $V^{-1/6}$. It is predicted 
that velocity of capillary blood is almost invariant ($\sim V^{\epsilon}$), 
capillary transit time scales as $V^{1/6}$, capillary length increases as 
$V^{1/6+\epsilon}$, and capillary number as $V^{2/3-\epsilon}$, where 
$\epsilon$ is typically a small correction for medium and large brains, 
due to blood viscosity dependence on capillary radius. 
It is shown that the amount of capillary length and blood flow per cortical 
neuron are essentially conserved across mammals. These results indicate 
that geometry and dynamics of global neuro-vascular coupling have 
a proportionate character. Moreover, cerebral metabolic, hemodynamic, 
and microvascular variables scale with allometric exponents that are 
simple multiples of 1/6, rather than 1/4, which suggests that brain 
metabolism is more similar to the metabolism of aerobic than resting body.
Relation of these findings to brain functional imaging studies involving 
the link between cerebral metabolism and blood flow is also discussed.
\end{abstract}

%\pacs{PACS Nos. 87.18.Sn, 87.19.La, 87.80.Xa, 89.40.+k, 87.23.Ge}
%\narrowtext 
%\maketitle 

%\begin{narrowtext}

%\maketitle

\newpage


\vspace{0.1cm}

\noindent Email: jkarbowski@duch.mimuw.edu.pl; jkarbowski@ibib.waw.pl

%\noindent Phone: 

\vspace{0.3cm}

\newpage

\noindent{\Large \bf Introduction}

It is well established empirically that whole body metabolism of resting 
mammals scales with body volume (or mass) with an exponent close to 3/4, 
which is known as Kleiber's law \cite{kleiber,schmidt,calder,dodds}. The same 
exponent or its simple derivatives govern the scalings of respiratory and 
cardiovascular systems in mammals and some other physiological parameters 
in animals and plants \cite{schmidt,calder,enquist}. Because of its almost 
ubiquitous presence, the quarter power has often been described as a general 
law governing metabolism and blood circulation, and several formal models 
explaining its origin have been proposed that still cause controversy 
\cite{west,banavar,darveau,savage}. However, as was found by the author 
\cite{karb2007}, the brain metabolism at rest seems to follow another scaling 
rule. Total brain metabolic rate (both oxygen and glucose) scales with 
brain volume with an exponent $\approx 0.85$, or close to 5/6 \cite{karb2007}. 
Consequently, the volume-specific cerebral metabolism decreases with brain 
size with an exponent around $-1/6$, and this value is highly homogeneous 
across many structures of gray matter \cite{karb2007}. The origin of these 
cerebral exponents has never been explained, although it is interesting why 
brain metabolism scales different than metabolism of other systems.

The brain, similar to other organs, uses capillaries for delivery of
metabolic nutrients (oxygen, glucose, etc.) to its cells \cite{krogh}.
Moreover, numerical density of cerebral capillaries is strongly correlated 
with brain hemodynamics and metabolism \cite{klein,borowsky}. However, the
cerebral microvascular network differs from other non-cerebral networks
in two important ways. First, in the brain there exists a unique physical
border, called the brain-blood barrier, which severely restricts influx
of undesired molecules and ions to the brain tissue. Second, cerebral
capillaries exhibit a large degree of physical plasticity, manifested in 
easy adaptation to abnormal physiological conditions. For instance, during 
ischemia (insufficient amount of oxygen in the brain) capillaries can
substantially modify their diameter to increase blood flow and hence oxygen
influx \cite{boero,hauck,ito}. These two factors, i.e. structural differences
and plasticity of microvessels, can in principle modify brain metabolism
in such a way to yield different scaling rules in comparison to e.g. lungs
or muscles. Another, related factor that may account for the uncommon 
brain metabolic scaling is the fact that brain is one of the most energy
expensive organs in the body \cite{karb2007,aiello}. This is usually 
attributed to the neurons with their extended axons and dendrites, which 
utilize relatively large amounts of glucose and ATP for synaptic communication
\cite{attwell,karb2009}.

The main purpose of this paper is to determine scaling laws for blood flow 
and geometry of capillaries in the brain of mammals. Are they different from
those found or predicted for cardiovascular and respiratory systems? If so, 
do these differences account for brain metabolic allometry? How the scalings 
of blood flow and capillary dimensions relate to the scalings of neural 
characteristics, such as neural density and axon (or dendrite) length? 
This study might have implications for expanding of our understanding of 
mammalian brain evolution, in particular the relationship between brain 
wiring, metabolism, and its underlying microvasculature 
\cite{karb2007,carmeliet,attwell2}. 
The results can also be relevant for research involving the microvascular 
basis of brain functional imaging studies, which use relationships between 
blood flow and metabolism to decipher regional neural activities 
\cite{heeger,logothetis}.

\newpage

\noindent{\Large \bf Results}

The data for brain circulatory system were collected from different sources
(see Materials and Methods). They cover several mammals spanning 3-4 orders 
of magnitude in brain volume, from mouse to human.

\vspace{0.4cm}

\noindent{\bf 1. Empirical scaling data.}

Cerebral blood flow CBF in different parts of mammalian gray matter decreases
systematically with gray matter volume, both in the cortical and subcortical
regions (Fig. 1). In the cerebral cortex, the scaling exponent for regional
CBF varies from $-0.13$ for the visual cortex (Fig. 1A), $-0.15$ for the
parietal cortex (Fig. 1B), $-0.17$ for the frontal cortex (Fig. 1C), to
$-0.19$ for the temporal cortex (Fig. 1D). The average cortical exponent
is $-0.16\pm 0.02$. In the subcortical regions, the CBF scaling exponent
is $-0.14$ for hippocampus (Fig. 2A), $-0.17$ for thalamus (Fig. 2B), and
$-0.18$ for cerebellum (Fig. 2C). The average subcortical exponent is identical
with the cortical one, i.e., $-0.16\pm 0.02$, and both of them are close to
$-1/6$. It is interesting to note that almost all of the cortical areas 
(except temporal cortex) have scaling exponents whose 95$\%$ confidence 
intervals do not include a quarter power exponent $-1/4$.

The microvessel system delivering energy to the brain consists of 
capillaries. The capillary diameter increases very weakly but significantly 
with brain size, with an exponent of 0.08 (Fig. 3A). On the contrary, the 
volume-density of capillary length decreases with brain size raised 
to a power of $-0.16$ (Fig. 3B). Thus, the cerebral capillary network 
becomes sparser as brain size increases. Despite this, the fraction of 
gray matter volume taken by capillaries is approximately independent of 
brain size (Fig. 3C). Another vascular characteristic, the arterial partial 
oxygen pressure, is also roughly invariant with respect to brain volume 
(Fig. 3D).

A degree of neurovascular coupling can be characterized by geometric  
relationships between densities of capillaries and neurons. Scaling of the 
density of neuron number in the cortical gray matter is not uniform across 
mammals \cite{houzel2006,houzel2007,houzel2011}. In fact, the scaling exponent 
depends to some extent on mammalian order and the animal sample used 
\cite{houzel2011}. For the sample of mammals used in this study, it is
found that cortical neuron density decreases with cortical gray matter volume 
with an exponent of $-0.13$ (Fig. 4A). This exponent is close to the 
exponent for the scaling of capillary length density, which is $-0.16$ 
(Fig. 3B). Consistent with that, the ratio of cortical capillary length 
density to neuron density across mammals is approximately constant and 
independent of brain size (Fig. 4B). Typically, there is about 10 $\mu$m
of capillaries per cortical neuron. The scaling dependence between the 
two densities yields an exponent close to unity (Fig. 4C), which shows 
a proportionality relation between them.

Cerebral blood flow CBF scales with brain volume the same way as does 
capillary length density (Figs. 1,2,3B), and thus, CBF should also be related 
to neural density. Indeed, in the cerebral cortex the ratio of the average 
CBF to cortical neural density is independent of brain scale (Fig. 5). 
This means that the average amount of cortical blood flow per neuron is 
invariant among mammals, and about $(1.45\pm 0.4)10^{-8}$ mL/min. Taken 
together, the findings in Figs. 4 and 5 suggest a tight global correlation 
between neurons and their energy supporting microvascular network.

\vspace{2.4cm}

\noindent{\bf 2. Theoretical scaling rules for cerebral capillaries.}

Below I derive theoretical predictions for the allometry of brain capillary
characteristics, such as: capillary length and radius, capillary number, 
blood velocity, and time taken by blood to travel through a capillary.
I also find relationships connecting cerebral metabolic rate and blood flow 
with neuron density. The following assumptions are made in the analysis: 
(i) Oxygen consumption rate in gray matter CMR$_{O2}$ scales with cortical
gray matter volume $V$ as $V^{-1/6}$, in accordance with Ref. \cite{karb2007};
(ii) Capillary volume fraction, $f_{c}= \pi N_{c}L_{c}R_{c}^{2}/V$, is
invariant with respect to $V$, which follows from the empirical results
in Fig. 3C. The symbol $N_{c}$ denotes total capillary number in the gray 
matter, $L_{c}$ is the length of a single capillary segment, and $R_{c}$
is its radius; (iii) Driving blood pressure $\Delta p_{c}$ through capillaries 
is independent of brain size, which is consistent with a known fact 
that arterial blood pressure (both systolic and diastolic) of resting mammals
is independent of body size \cite{woodbury,gregg,li}; 
(iv) Partial oxygen pressure $p_{O2}$ in capillaries is also invariant,
which is consistent with the empirical data in Fig. 3D on the invariance
of arterial oxygen pressure;
(v) Cerebral blood flow CBF is proportional to oxygen consumption rate
CMR$_{O2}$, due to adaptation of capillary diameters to oxygen demand.

The cerebral metabolic rate of oxygen consumption CMR$_{O2}$, according
to the modified Krogh model \cite{krogh,boero}, is proportional to the
product of oxygen flux through capillary wall and the tissue-capillary
gradient of oxygen pressure $\Delta p_{O2}$, i.e.

\begin{equation}
\mbox{CMR}_{O2}\sim D\left(\frac{N_{c}L_{c}}{V}\right) \Delta p_{O2},  
\end{equation}\\
where $D$ is the oxygen diffusion constant in the brain. The dependence
of CMR$_{O2}$ on capillary radius in this model has mainly a logarithmic 
character, and hence it is neglected as weak. Since oxygen pressure in the 
brain tissue is very low \cite{lenigert}, the pressure gradient 
$\Delta p_{O2}$ is essentially equal to the capillary oxygen pressure 
$p_{O2}$. Consequently, the formula for CMR$_{O2}$ simplifies to 
$\mbox{CMR}_{O2} \sim \rho_{c} p_{O2}$, where $\rho_{c}$ is the density
of capillary length $\rho_{c}= N_{c}L_{c}/V$.

From the assumptions (i) and (iv) we obtain that capillary length density
$\rho_{c} \sim V^{-1/6}$. Additionally, from (ii) we have 
$R_{c}^{2} \sim f_{c}/\rho_{c} \sim V^{0}/V^{-1/6} \sim V^{1/6}$, implying
that capillary radius (or diameter) $R_{c}$ scales as $V^{1/12}$.
Consequently capillary diameter does not increase much with brain 
magnitude. As an example, a predicted capillary diameter for elephant with its
cortical gray matter volume 1379 cm$^{3}$ \cite{hakeem} is 7.2 $\mu$m, which 
does not differ much from those of rat (4.1 $\mu$m \cite{hauck,michaloudi}) 
or human (6.4 $\mu$m \cite{meier,lauwers}), who have corresponding volumes
3450 and 2.4 times smaller.

The blood flow $Q_{c}$ through a capillary is governed by a modified
Poiseuille's law in which blood viscosity depends on capillary radius  
\cite{sugihara}:

\begin{equation}
Q_{c}= \frac{\pi \Delta p_{c} R_{c}^{4}}{8\eta_{ef}(R_{c}) L_{c}},  
\end{equation}\\
where $\Delta p_{c}$ is the axial driving blood pressure along a capillary
of length $L_{c}$, and $\eta_{ef}(R_{c})$ is the capillary radius dependent 
effective blood viscosity. The latter dependence has a nonmonotonic character,
i.e. for small diameters the viscosity $\eta_{ef}(R_{c})$ initially decreases 
with increasing $R_{c}$, reaching a minimum at diameters about $5-7$ $\mu$m. 
For $2R_{c} > 10$ $\mu$m the blood viscosity $\eta_{ef}$ slowly increases 
with $R_{c}$ approaching its bulk value for diameters $\sim$ 500 $\mu$m. 
This phenomenon is known as the Fahraeus-Lindqvist effect \cite{fahraeus}. 
In general, blood viscosity in narrow microvessels depends on microvessel 
thickness because red blood cells tend to deform and place near the center 
of capillary leaving a cell-free layer near the wall \cite{sugihara,pries}. 
These two regions have significantly different viscosities, with the 
cell-free layer having essentially plasma viscosity $\eta_{p}$, which is much 
smaller than the bulk (or center region) viscosity $\eta_{c}$. The formula 
relating the effective blood viscosity $\eta_{ef}$ with capillary radius 
and both viscosities $\eta_{p}$ and $\eta_{c}$ is given by \cite{sugihara}:

\begin{equation}
\eta_{ef}(R_{c})= \frac{\eta_{p}}{1-(1-\mu)(1-w/R_{c})^{4}},  
\end{equation}\\
where $\mu= \eta_{p}/\eta_{c} \ll 1$, and $w$ is the thickness of cell-free
layer.

For capillary radiuses relevant for the brain, i.e. 1.5 $\mu$m $ < R_{c} < $ 
3.5 $\mu$m (see Suppl. Table S2), the ratio $w/R_{c}$ increases with increasing
$R_{c}$, which causes a decline in the effective blood viscosity down to
its minimal value at $R_{c}=3-3.5$ $\mu$m (Table 1). Using the data in
Table 1 taken from \cite{sugihara}, we can approximate the denominator in 
Eq. (3) for this range of radiuses by a simple, explicit function of $R_{c}$. 
The best fit is achieved with a logarithmic function, i.e.
$1-(1-\mu)(1-w/R_{c})^{4} \approx 0.85[\ln(R_{c}/R_{o})]^{2/3}$, where
$R_{o}= 1.2$ $\mu$m (Table 1). As a result, the effective blood viscosity 
takes a simple form:

\begin{equation}
\eta_{ef}(R_{c})= 1.18\eta_{p} \left(\ln(R_{c}/R_{o})\right)^{-2/3}.  
\end{equation}\\

Cerebral blood flow CBF in the brain gray matter is defined as 
$\mbox{CBF}= Q/V$, where $Q= N_{c}Q_{c}$ is the total capillary blood
flow through all $N_{c}$ capillaries. Thus CBF is given by

\begin{equation}
\mbox{CBF} \sim \frac{\Delta p_{c}}{\eta_{ef}(R_{c})} 
\frac{\rho_{c}R_{c}^{4}}{L_{c}^{2}},  
\end{equation}\\
or
\begin{equation}
\mbox{CBF} \sim \frac{\Delta p_{c}\rho_{c}R_{c}^{4}}
{\eta_{p}L_{c}^{2}}\left[\ln\left(\frac{R_{c}}{R_{o}}\right)\right]^{2/3}. 
\end{equation}\\
We can rewrite the logarithm present in Eq. (6), in an equivalent form, 
as a power function $(R_{c}/R_{o})^{\gamma}$ with a variable exponent
$\gamma$ given by (see Appendix S1 in the Supp. Infor.):

\begin{equation}
\gamma= \frac{2}{3}\frac{\ln(\ln(R_{c}/R_{o}))}{\ln(R_{c}/R_{o})}, 
\end{equation}\\
so that CBF becomes

\begin{equation}
\mbox{CBF} \sim \frac{\Delta p_{c}}{\eta_{p}} 
\frac{\rho_{c}R_{c}^{4+\gamma}}{L_{c}^{2}}.  
\end{equation}\\
The exponent $\gamma$ in this equation can be viewed as a correction due
to non-constant blood viscosity (Fahraeus-Lindqvist effect \cite{fahraeus}).
The dependence of $\gamma$ on the capillary diameter is shown in Table 2.
Because in general $\gamma < 0$, its presence in Eq. (8) reduces the power
of $R_{c}$. However, this effect is weak for medium and large brains as
$|\gamma| \ll 1$. Even for a small rat brain the relative influence of
$\gamma$ is rather weak, since $|\gamma|/4 \approx 0.19$. In contrast,
for very small brains, such as mouse, the effect caused by $\gamma$ is
strong (Table 2), which reflects a sharp increase in the effective blood
viscosity for the smallest capillaries \cite{sugihara,pries}.

Now we are in a position to derive scaling rules for the capillary length
segment $L_{c}$, capillary blood velocity $u_{c}$, and the number of
capillaries $N_{c}$. From Eq. (8), using the assumptions (i), (iii), and (v),
we obtain $L_{c}^{2} \sim  \rho_{c}R_{c}^{4+\gamma}/\mbox{CMR}_{O2}$,
which implies that $L_{c} \sim R_{c}^{2+\gamma/2}$ (viscosity of blood
plasma is presumably independent of brain scale \cite{amin}). Consequently,
$L_{c} \sim V^{1/6+\gamma/24}$, i.e. capillary length should weakly increase
with brain size. Although there are no reliable data on $L_{c}$, we can 
compare our prediction with the measured intercapillary distances, which 
generally should be positively correlated with $L_{c}$. Indeed, the mean 
intercapillary distance in gray matter increases with increasing brain volume, 
and is $17-24$ $\mu$m in rat \cite{schlageter}, 24 $\mu$m in cat \cite{pawlik}, 
and 58 $\mu$m in human \cite{meier}.

Average velocity $u_{c}$ of blood flow in brain capillaries is given by
$u_{c}= Q_{c}/(\pi R_{c}^{2})$. Using the expressions for $Q_{c}$ and 
$\eta_{ef}$, we get $u_{c} \sim (\Delta p_{c}R_{c}^{2+\gamma})/L_{c}$. Since 
$L_{c} \sim R_{c}^{2+\gamma/2}$ above, and using the assumption (iii), we
obtain $u_{c} \sim R_{c}^{\gamma/2} \sim V^{\gamma/24}$. Thus, capillary
blood velocity is almost independent of brain size for medium and large
brains, as then $\gamma \rightarrow 0$ (Table 2). For very small brains,
instead, there might be a weak dependence. A related quantity, the blood 
transit time $\tau_{c}$ through a capillary, defined as 
$\tau_{c}= L_{c}/u_{c}$, scales as $\tau_{c} \sim V^{1/6}$, regardless
of the brain magnitude. This indicates that $\tau_{c}$ and CBF are inversely 
related across different species, $\tau_{c} \sim \mbox{CBF}^{-1}$, because
of their scaling properties.

We can find the scaling relation for the total number of capillaries $N_{c}$
from the volume-density of capillary length $\rho_{c}$. We obtain
$N_{c}= \rho_{c}V/L_{c} \sim V^{-1/6}V/V^{1/6+\gamma/24} 
\sim V^{2/3-\gamma/24}$, i.e. the exponent for $N_{c}$ is close to 2/3
for not too small brains. As an example, the number of 
capillary segments in the human cortical gray matter should be 123 times 
greater than that in the rat (cortical volumes of both hemispheres in 
rat and human are 0.42 cm$^{3}$ \cite{houzel2006} and 572.0 cm$^{3}$ 
\cite{houzel2010}, respectively).

As was shown above, CMR$_{O2}$ must be proportional to the volume density of 
capillary length $\rho_{c}$ (Eq. 1). On the other hand, the empirical results 
in Fig. 4 indicate that $\rho_{c}$ is roughly proportional to neuron density 
$\rho_{n}$. Thus, we have approximately CMR$_{O2}$ $\sim \rho_{n}$ across 
different mammals. This implies that oxygen metabolic energy per neuron in 
the gray matter should be approximately independent of brain size. Exactly 
the same conclusion was reached before in a study by Herculano-Houzel 
\cite{houzel2011}, based on independent data analysis. Moreover, since 
cortical CMR$_{O2}$ and CBF scale the same way against brain size, we also 
have $\mbox{CBF} \sim \rho_{n}$, which is confirmed by the results in Fig. 5. 
In other words, both cerebral metabolic rate and blood flow per neuron are 
scale invariant.

\newpage

\noindent{\Large \bf  Discussion}

\vspace{0.4cm}

\noindent{\bf 1. General discussion.}

The summary of the scaling results is presented in Table 3. Some of these
allometric relations are directly derived from the experimental data
(CBF, $R_{c}$, $\rho_{c}$, $f_{c}$, $\rho_{c}/\rho_{n}$, CBF/$\rho_{n}$), 
and others are theoretically deduced 
($N_{c}$, $L_{c}$, $u_{c}$, $\tau_{c}$). The interesting result is that 
cerebral blood flow CBF in gray matter scales with cortical gray matter volume 
raised to a power of $-0.16$. The similar exponent governs the allometry of 
cortical metabolic rate CMR \cite{karb2007}, which indicates that brain 
metabolism and blood flow are roughly linearly proportional across different 
mammals. This conclusion is compatible with several published studies that 
have shown the proportionality of CMR and CBF on a level of a single animal 
(rat, human) across different brain regions \cite{klein,roland}.

The coupling between CMR and CBF manifests itself also in their relation to 
the number of neurons. In this respect, the present study extends the recent 
result of Herculano-Houzel \cite{houzel2011} about the constancy of metabolic 
energy per neuron in the brains of mammals, by showing that also cerebral 
blood flow and capillary length per neuron are essentially conserved across 
species. There are approximately 10 $\mu$m of capillaries and 
$1.45\cdot 10^{-8}$ mL/min of blood flow per cortical neuron
(Figs. 4 and 5; Supp. Tables S2 and S3). This finding suggests that 
not only brain metabolism but also its hemodynamics and microvascularization 
are evolutionarily constrained by the number of neurons. This mutual coupling 
might be a result of optimization in the design of cerebral 
energy expenditure and blood circulation.

It should be underlined that both CBF and CMR scale with brain volume with
the exponent about $-1/6$, which is significantly different from the exponent 
$-1/4$ relating whole body resting specific metabolism with body 
volume  \cite{kleiber,schmidt,calder}. Instead, the cerebral exponent
$-1/6$ is closer to an exponent $-0.12\pm 0.02$ characterizing maximal body 
specific metabolic rate and specific cardiac output in strenuous exercise
\cite{bishop99,weibel2005}. In this sense, the brain metabolism and 
its hemodynamics resemble more the metabolism and circulation of exercised
muscles than other resting organs, which is in line with the empirical
evidence that brain is an energy expensive organ 
\cite{karb2007,aiello,attwell}. This may also suggest that there exists a 
common plan for the design of microcirculatory system in different parts of 
the mammalian body that uses the same optimization principles 
\cite{weibel91}.

The results of this study show that as brain increases in size its capillary 
network becomes less dense, i.e. the densities of both capillary number and 
length decrease, respectively as $N_{c}/V \sim V^{-1/3-\gamma/24}$ and 
$\rho_{c} \sim V^{-1/6}$ (Table 3). Contrary to that, the capillary dimensions 
increase weakly with brain volume, their radius as $R_{c} \sim V^{1/12}$ and 
their length segment as $L_{c} \sim V^{1/6+\gamma/24}$, which are sufficient to 
make the fraction $f_{c}$ of capillary volume in the gray matter to be scale 
invariant (Table 3). The correction $\gamma/24$ appearing in the scaling
exponents for $N_{c}/V$ and $L_{c}$ reflects the fact that blood viscosity
depends on capillary radius (Fahraeus-Lindqvist effect \cite{fahraeus}).
This correction is however small for sufficiently large brains, generally
for brains larger or equal to that of rat, for which typical values of
$\gamma/24$ are in the range from $-0.032$ to $-0.0004$ (Table 2). 
On the contrary, for brains of mouse size or smaller, this correction is 
substantial, about $-0.15$, which implies that for very small brains $L_{c}$
is essentially constant.

Despite the changes in the geometry of microvessels, the velocity of capillary 
blood $u_{c}$ is almost scale invariant for not too small brains (exponent
$\gamma/24\approx 0$; Table 3). This prediction agrees with direct measurements 
of velocity in the brains of mouse, rat, and cat, which does not seem to change
much, i.e. it is in the range $1.5-2.2$ mm/sec \cite{pawlik,unekawa}. 
Consequently the transit time $\tau_{c}$ through a capillary increases with 
brain size as $\tau_{c} \sim V^{1/6}$, i.e. the scaling exponent is again $1/6$. 
Another variable that seems to be independent of brain scale is partial oxygen 
pressure in cerebral capillaries (Table 3), which is consistent with the 
empirical findings in Fig. 3D on the invariance of oxygen pressure in arteries, 
as the two circulatory systems are mutually interconnected.

\vspace{0.4cm}

\noindent{\bf 2. Capillary scaling in cerebral and non-cerebral tissue.}

The above scaling results for the brain can be compared with available 
analogous scaling rules for pulmonary, cardiovascular, and muscle systems. 
For these systems, it was proposed (no direct measurements) that partial oxygen
pressure in capillaries should decline weakly with whole body volume (or organ 
volume as lung and heart volumes, $V_{lung}, V_{heart}$, scale isometrically 
with body volume \cite{schmidt}) with an exponent around $-1/12$, to account 
for the whole body specific metabolic exponent $-1/4$  \cite{dawson,west2}. 
In the resting pulmonary system, the capillary radius as well as the density 
of capillary length scale the same way as they do in the brain, i.e., with 
the exponents 1/12 and $-1/6$, respectively, against system's volume 
\cite{dawson2008}. Also, the capillary blood velocity in cerebral and 
non-cerebral tissues scale similarly, at least for not too small volumes,
i.e. both are scale invariant \cite{schmidt,calder} (Table 3). However, 
the number of capillaries and capillary length seem to scale slightly 
different in the resting lungs, i.e. $N_{c} \sim V_{lung}^{5/8}$ and 
$L_{c} \sim V_{lung}^{5/24}$ \cite{dawson}, although the difference can
be very mild. For the resting heart, it was predicted (again, no direct 
measurements) that $N_{c} \sim V_{heart}^{3/4}$, and blood transit time 
through a capillary $\tau \sim V_{heart}^{1/4}$ \cite{west2}, i.e. the 
exponents are multiples of a quarter power and are slightly larger than those
for the brain (Table 3). Interestingly, for muscles and lungs in mammals 
exercising at their aerobic maxima, the blood transit time scales against 
body mass with an exponent close to 1/6 \cite{kayar}, which is the same as 
in the brain (Table 3). This again suggests that brain metabolism is similar 
to the metabolism of other maximally exercised organs. Overall, the small 
differences in the capillary characteristics among cerebral and non-cerebral 
resting tissues might account for the observed differences in the 
allometries of brain metabolism and whole body resting metabolism. 
In particular, the prevailing exponent 1/6 found in this study for brain 
capillaries, instead of 1/4, seems to be a direct cause for the distinctive 
brain metabolic scaling.

\vspace{0.4cm}

\noindent{\bf 3. Brain microvascular network vs. neural network.}

The interesting question from an evolutionary perspective is how the 
allometric scalings for brain capillary dimensions relate to the allometry 
of neural characteristics. The neural density $\rho_{n}$ (number of cortical 
neurons $N_{n}$ per cortical gray matter volume $V$) scales with cortical 
volume with a similar exponent as does the density of capillary length 
$\rho_{c}$ (Fig. 4A). Thus, as a coarse-grained global description we have 
approximately $\rho_{n} \sim \rho_{c}$ (Fig. 4B,C), or 
$N_{n} \sim N_{c}L_{c}$. 
The latter relation means that the total number of neurons is roughly 
proportional to the total length of capillaries, or equivalently, that 
capillary length per cortical neuron is conserved across different mammals.
This cross-species conclusion is also in agreement with the experimental data
for a single species. In particular, for mouse cerebral cortex it was found 
that densities of neural number and microvessel length are correlated globally 
across cortical areas (but not locally within a single column) \cite{tsai}.
Moreover, since axons and dendrites occupy a constant fraction of cortical 
gray matter volume (roughly 1/3 each; \cite{braitenberg,chklovskii}), we have 
$N_{n}ld^{2} \sim V$, where $l$ and $d$ are respectively axon (or dendrite)
length per neuron and diameter. Furthermore, because the average axon diameter
$d$ (unmyelinated) in the cortical gray matter is approximately invariant 
against the change of brain scale \cite{braitenberg,olivares}, we obtain
the following chain of proportionalities:
$l \sim \rho_{n}^{-1} \sim \rho_{c}^{-1} \sim V^{1/6} \sim L_{c}^{\alpha}$,
where the exponent $\alpha= 1/(1+\gamma/4)$. For medium and large brains, 
$\alpha \approx 1$, implying a nearly proportional dependence of axonal and 
dendritic lengths on capillary segment length. For very small brains (roughly
below the volume of rat brain), $\alpha$ can be substantially greater than
1, suggesting a non-linear dependence between capillary and neural sizes.

Given that the main exchange of oxygen between blood and brain takes place 
in the capillaries, these results suggest that metabolic needs of larger
brains with greater but numerically sparser neurons must be matched by 
appropriately longer yet sparser capillaries. This finding reflects a rough,
global relationship, which might or might not be related to the fact that 
during development neural and microvessel wirings share mutual mechanisms 
\cite{carmeliet,stubbs}. At the cortical microscale, however, things could be 
more complicated, and a neuro-vascular correlation might be weaker, as both 
systems are highly plastic even in the adult brain (e.g. \cite{chklovskii2}). 
Regardless of its nature and precise dependence, the neuro-vascular coupling 
might be important for optimization of neural wiring 
\cite{chklovskii,mitchison,karb2001}. In fact, neural connectivity in the 
cerebral cortex is very low, and it decreases with brain size 
\cite{karb2001,karb2003}, similar to the density of capillary length
(Fig. 3B, Table 3). To make the neural connectivity denser, it would require 
longer axons and consequently longer capillaries. That may in turn increase 
excessively brain volume and its energy consumption, i.e. the costs of brain 
maintenance. As a result, the metabolic cost of having more neural 
connections and synapses for storing memories might outweigh its functional 
benefit.

The brain metabolism is obviously strictly related to neural activities.
In general, higher neural firing rates imply more cerebral energy consumed
\cite{attwell,karb2009}. It was estimated, based on a theoretical formula 
relating CMR with firing rate, that the latter should decline with brain size 
with an exponent around $-0.15$ \cite{karb2009}. This implies that neurons 
in larger brain are on average less active than neurons in smaller brains. 
Such sparse neural representations may be advantageous in terms of saving 
the metabolic energy \cite{attwell,levy,laughlin}. At the same time, what 
may be related, neural activity is distributed in such a way that both the 
average energy per neuron and the average blood flow per neuron are 
approximately invariant with respect to brain size 
(Fig. 5; Table 3, \cite{houzel2011}). Additionally, average firing rate 
should be inversely proportional to the average blood transit time $\tau_{c}$ 
through a capillary, because both of them scale reversely with brain size 
(Table 3). Thus, it appears that global timing in neural activities should 
be correlated with the timing of cerebral blood flow. These general 
considerations suggest that apart from structural neuro-vascular coupling 
there is probably also a significant dynamic coupling. This conclusion 
is qualitatively compatible with experimental observations in which enhanced 
neural activity is invariably accompanied by increase in local blood flow 
\cite{moore}.

\vspace{0.4cm}

\noindent{\bf 4. Relationship to brain functional imaging.}

The interdependencies between brain metabolism, blood flow, and capillary 
parameters can have practical meaning. Currently existing techniques for
non-invasive visualization of brain function, such as PET or fMRI, are
associated with measurements of blood flow CBF and oxygen consumption 
CMR$_{O2}$. It turns out that during stimulation of a specific brain region, 
CBF increases often, but not always, far more than CMR$_{O2}$ \cite{raichle}.
However, both of them increase only by a small fraction in relation to
the background activity, even for massive stimulation \cite{moore,raichle}.
This phenomenon was initially interpreted as an uncoupling between blood
perfusion and oxidative metabolism \cite{fox86}. Later, it was shown that
this asymmetry between CBF and CMR$_{O2}$ can be explained in terms of
mechanistic limitations on oxygen delivery to brain tissue through blood
flow \cite{buxton}. We can provide a related, but simpler explanation of 
these observations that involves physical limitations on the relative 
changes in capillary oxygen pressure and radius.

During brain stimulation, both CBF and CMR$_{O2}$ change by $\delta\mbox{CBF}$
and $\delta\mbox{CMR}_{O2}$, which are according to Eqs. (1) and (8) related
to modifications in capillary radius (from $R_{c}$ to $R_{c}+\delta R_{c}$),
and changes in partial oxygen pressure ($p_{O2} \mapsto p_{O2}+\delta p_{O2}$). 
The density of perfused capillary length $\rho_{c}$ remains constant for normal 
neurophysiological conditions. Accordingly, a small fraction of 
blood flow change is

\begin{equation}
\frac{\delta\mbox{CBF}}{\mbox{CBF}} \approx (4+\gamma)\frac{\delta R_{c}}{R_{c}} 
\end{equation}\\
and similarly, a small fractional change in the oxygen metabolic rate is:

\begin{equation}
\frac{\delta\mbox{CMR}_{O2}}{\mbox{CMR}_{O2}} \approx \frac{\delta p_{O2}}{p_{O2}}.
\end{equation}\\
In general, oxygen pressure increases with increasing capillary radius,
in response to increase in blood flow CBF. This relationship can have a
complicated character. We simply assume that $p_{O2} \sim R_{c}^{a}$,
where the unknown exponent $a$ $(a > 0)$ contains all the non-linear effects,
however complicated they are. Thus, a small fractional change in oxygen
pressure can be written as $\delta p_{O2}/p_{O2} \approx a \delta R_{c}/R_{c}$.
As a result, we obtain 

\begin{equation}
\frac{\delta\mbox{CMR}_{O2}}{\mbox{CMR}_{O2}} \approx 
\frac{a}{(4+\gamma)} \frac{\delta\mbox{CBF}}{\mbox{CBF}}.
\end{equation}\\
If partial oxygen pressure $p_{O2}$ depends on capillary radius linearly or 
sublinearly, i.e., if $a \le 1$, then  the fractional increase in oxygen 
metabolism is significantly smaller than a corresponding increase in cerebral 
blood flow. This case corresponds to the experimental reports showing that 
this ratio is $\ll 1$, for example, in the visual cortex ($\sim 0.1$) 
\cite{fox88} and in the sensory cortex ($\sim 0.2-0.4$)  \cite{fox86,seitz}. 
If, in turn, $p_{O2}$ depends on $R_{c}$ superlinearly, i.e. if $a > 1$, then
the coefficient $a/(4+\gamma)$ in Eq. (10) can be of the order of unity.
Such cases have been also reported experimentally during
cognitive activities \cite{roland} or anesthesia \cite{nilsson,smith}.

\newpage

\noindent{\Large \bf  Materials and Methods}

The ethics statement does not apply to this study.
CBF data were collected from different sources: for mouse \cite{frietsch2007}, 
rat \cite{frietsch2000}, rabbit \cite{tuor}, cynomolgus monkey \cite{orlandi}, 
rhesus monkey \cite{noda}, pig \cite{delp}, and human \cite{bentourkia}. 
Cerebral capillary characteristics were obtained from several sources: 
for mouse \cite{boero,tsai}, 
rat \cite{hauck,bar,michaloudi}, cat \cite{pawlik,tieman}, dog 
\cite{luciano}, rhesus monkey \cite{weber}, and human \cite{meier,lauwers}. 
Data for calculating neuron densities were taken from 
\cite{houzel2006,houzel2007,houzel2010,braitenberg,mayhew,haug}.
Cortical volume data (for 2 hemispheres) are taken from 
\cite{braitenberg,houzel2010,mayhew}. Their values are: mouse  0.12 cm$^{3}$, 
rat 0.42 cm$^{3}$, rabbit 4.0 cm$^{3}$, cat 14.0 cm$^{3}$, cynomolgus monkey  
21.0 cm$^{3}$, dog  35.0 cm$^{3}$, rhesus monkey 42.9 cm$^{3}$, 
pig 45.0 cm$^{3}$, human 571.8 cm$^{3}$. 
All the numerical data are provided in the Supporting Information
(Tables S1, S2, and S3).

\newpage

\noindent{\Large \bf  Supporting Information}

\vspace{0.3cm}

\noindent{\bf Appendix S1}

\vspace{0.3cm}

\noindent{\bf Table S1} \\
Regional cerebral blood flow CBF in mammals.

\vspace{0.3cm}

\noindent{\bf Table S2} \\
Cerebral capillary and neural characteristics in mammals.

\vspace{0.3cm}

\noindent{\bf Table S3} \\
Arterial partial oxygen pressure and average cortical CBF per neuron.

\vspace{1.5cm}

%\noindent{\bf Acknowledgments}
%
%The work was supported by the grant from the Polish Ministry of Science
%and Education (NN 518 409238), and by the Marie Curie Actions EU grant 
%FP7-PEOPLE-2007-IRG-210538.  

\newpage

\vspace{1.5cm}

%\noindent{\bf\large  References} \\

%\vspace{1.5cm}

\newpage

{\bf \large Figure Captions}

Fig. 1\\
Scaling of cerebral blood flow CBF in the cortical gray matter. 
(A) Visual cortex: $y=-0.127x+1.98$ ($R^{2}=0.93$, $p < 0.001$).
95$\%$ confidence interval for the slope CI=(-0.168,-0.086). 
(B) Parietal cortex: $y=-0.150x+2.00$ ($R^{2}=0.89$, $p=0.005$), slope
CI=(-0.222,-0.078).
(C) Frontal cortex: $y=-0.170x+2.00$ ($R^{2}=0.89$, $p=0.002$), slope
CI=(-0.239,-0.100).
(D) Temporal cortex: $y=-0.191x+2.09$ ($R^{2}=0.89$, $p=0.005$), slope
CI=(-0.286,-0.096).

\vspace{0.3cm}

Fig. 2\\
Scaling of cerebral blood flow CBF in the subcortical gray matter. 
(A) Hippocampus:  $y=-0.135x+1.89$ ($R^{2}=0.90$, $p=0.049$), 
slope CI=(-0.271,0.000).
(B) Thalamus: $y=-0.167x+2.02$ ($R^{2}=0.83$, $p=0.011$),
slope CI=(-0.272,-0.062). 
(C) Cerebellum: $y=-0.177x+2.03$ ($R^{2}=0.88$, $p=0.002$),
slope CI=(-0.252,-0.102). 

\vspace{0.3cm}

Fig. 3\\
Scaling of brain capillary characteristics against brain size. 
(A) Capillary diameter scales against cortical gray matter volume with 
the exponent 0.075 ($y=0.075x+0.60$, $R^{2}=0.87$, $p=0.007$), 
exponent CI=(0.034,0.117).
(B) Volume density of capillary length $\rho_{c}$ scales with the exponent 
$-0.16$ ($y=-0.162x+2.86$, $R^{2}=0.79$, $p=0.044$), exponent CI=(-0.316,-0.008).
(C) Fraction of capillary volume $f_{c}$ in gray matter is essentially independent 
of brain size ($y=0.029x-1.83$, $R^{2}=0.07$, $p=0.662$), the same as
(D) the arterial partial oxygen pressure 
($y=0.012x+1.96$, $R^{2}=0.09$, $p=0.472$).

\vspace{0.3cm}

Fig. 4\\
Neuron density versus capillary length density in the cerebral cortex. 
(A) Across our sample of mammals, the cortical neuron number density $\rho_{n}$ 
scales against cortical volume with the exponent $-0.13$ 
($y=-0.128x+1.86$, $R^{2}=0.87$, $p=0.022$), exponent CI=(-0.221,-0.036). 
(B) The ratio of the density of capillary length $\rho_{c}$ to the density 
of neurons $\rho_{n}$ in the cortex does not correlate with brain size 
($y=-0.034x+0.99$, $R^{2}=0.09$, $p=0.617$), exponent CI=(-0.228,0.161). 
(C) The log-log dependence of the capillary length density $\rho_{c}$ on neuron 
density $\rho_{n}$ gives the exponent of 1.05 
($y=1.051x+0.88$, $R^{2}=0.63$, $p=0.109$).

\vspace{0.3cm}

Fig. 5\\
Invariance of cerebral blood flow per cortical neuron across mammals. 
The ratio of CBF to neuron density $\rho_{n}$ in the cerebral cortex does not 
correlate significantly with brain size (log-log plot yields $y=-0.033x+0.183$, 
$R^{2}=0.39$, $p=0.261$). The value of CBF for each species is the arithmetic
mean of regional CBF across cerebral cortex.

\newpage

\begin{table}
\begin{center}
\caption{Parameters affecting the effective blood viscosity.}
\begin{tabular}{|l l l l l|}
\hline
\hline
$R_{c}$ [$\mu$m] & $w$ [$\mu$m] & $w/R_{c}$ & $1-(1-\mu)(1-w/R_{c})^{4}$
&  $0.85(\ln(R_{c}/1.2))^{2/3}$  \\

\hline
\hline

1.5  &  0.07  &  0.05  &  0.27  & 0.31  \\
2.0  &  0.30  &  0.15  &  0.54  & 0.54  \\  
2.5  &  0.60  &  0.24  &  0.71  & 0.69  \\
3.0  &  0.90  &  0.30  &  0.79  & 0.80  \\

\hline

\hline
\end{tabular}
\end{center}
Data for $w$ and $R_{c}$ were collected from \cite{sugihara}. The value
of $\mu$ was taken as 1/8. The last column represents values of the fitting
function to the function in the fourth column.
\end{table}

%\newpage

\begin{table}
\begin{center}
\caption{Exponent $\gamma$ as a function of capillary diameter.}
\begin{tabular}{|l l l l l l l|}
\hline
\hline
Species & mouse & rat & cat & dog & monkey & human  \\

\hline
\hline

$2R_{c}$ [$\mu$m]  & 3.1  & 4.1  &  5.1 &  4.5  &  5.6  &  6.4  \\
$\gamma$  &  -3.51  & -0.77  &  -0.25  &  -0.49  & -0.13  & -0.01 \\  

\hline

\hline
\end{tabular}
\end{center}
Data for $R_{c}$ were taken from Suppl. Inform. Table S2 (references therein).
\end{table}

%\newpage

\begin{table}
\begin{center}
\caption{Summary of scalings for brain capillaries and hemodynamics 
against cortical gray matter volume $V$.}
\begin{tabular}{|l l|}
\hline
\hline
Parameter &   Scaling rule  \\

\hline
\hline

Capillary radius, $R_{c}$           &      $R_{c} \sim V^{1/12}$   \\
Capillary length density, $\rho_{c}$    &      $\rho_{c} \sim V^{-1/6}$   \\
Capillary volume fraction, $f_{c}$        &      $f_{c} \sim V^{0}$   \\
Total capillary number, $N_{c}$         &   $N_{c} \sim V^{2/3-\gamma/24}$  \\
Capillary segment length, $L_{c}$       &  $L_{c} \sim V^{1/6+\gamma/24}$  \\
Capillary blood velocity, $u_{c}$       &    $u_{c} \sim V^{\gamma/24}$   \\
Capillary transit time, $\tau_{c}$      &     $\tau_{c} \sim V^{1/6}$   \\
Capillary oxygen pressure, $p_{O2}$      &      $p_{O2} \sim V^{0}$   \\
Capillary length per neuron, $\rho_{c}/\rho_{n}$    &  
$\rho_{c}/\rho_{n} \sim V^{0}$   \\
Cerebral blood flow, $\mbox{CBF}$   &      $\mbox{CBF} \sim V^{-1/6}$   \\
Blood flow per neuron, $\mbox{CBF}/\rho_{n}$   &      
$\mbox{CBF}/\rho_{n} \sim V^{0}$  \\
Oxygen consumption rate, $\mbox{CMR}_{O2}$   &      
$\mbox{CMR}_{O2} \sim V^{-1/6}$  \\
Oxygen use per neuron, $\mbox{CMR}_{O2}/\rho_{n}$   &      
$\mbox{CMR}_{O2}/\rho_{n} \sim V^{0}$  \\

\hline

\hline
\end{tabular}
\end{center}
\end{table}

%\newpage

\newpage

%\begin{table}
\begin{center}
%\caption{
{\bf Table S1:} Regional cerebral blood flow CBF in mammals.
\begin{tabular}{|l l l l l l l l|}
\hline
\hline
Species & visual  & frontal & temporal & parietal & hippocam. & thalamus & cerebell.  \\
        & cortex &  cortex  & cortex  &  cortex   &             &         &           \\

\hline
\hline

Mouse $^{(a)}$ & 124.0  &  165.0  &  157.0  &  166.0  &  112.8  &  195.5  &  195.0   \\ 
Rat  $^{(b)}$  & 116.0  &  124.0  &  195.0  &  113.0  &  77.8   &  116.0  &  123.0   \\
Rabbit $^{(c)}$ & 69.6  &   67.2  &  ---    &  63.6   &  ---    &  60.6   &  62.4   \\
Cynomolgus     &        &         &         &         &         &         &       \\
monkey $^{(d)}$ &  55.4  &  48.7  &   70.6  &  56.8  &   ---  &   56.7 &  57.2 \\
Rhesus monkey $^{(e)}$ &   59.2  &  45.0  &  53.2 & --- & 53.9 & ---    &  47.8  \\
Pig $^{(f)}$  & 66.5    &  63.3  &   45.5  &  62.8   &  41.0    &  48.5   &  64.0   \\
Human $^{(g)}$  &  43.0  &  41.2  &  44.6  &   43.4  &  ---  &   47.5  &  41.5   \\

\hline

\hline
\end{tabular}
\end{center} 
CBF data are given in mL/(100g*min). \\ 
References:
$^{(a)}$  \cite{frietsch2007};
$^{(b)}$  \cite{frietsch2000};
$^{(c)}$  \cite{tuor};
$^{(d)}$  \cite{orlandi}; 
$^{(e)}$  \cite{noda};
$^{(f)}$  \cite{delp}; 
$^{(g)}$  \cite{bentourkia}.

%\end{table}

%\newpage

%\begin{table}
\begin{center}
%\caption{
{\bf Table S2:} Cerebral capillary and neural characteristics in mammals.
\begin{tabular}{| l l l l l l |}
\hline
\hline
Species  &   $2R_{c}$ & $f_{c}$ &  $\rho_{c}$  & $\rho_{n}$ & $\rho_{c}/\rho_{n}$ 
  \\
     & [$\mu$m]  &    &  [mm/mm$^{3}$]  &  [10$^{6}$/cm$^{3}$] & [$\mu$m] \\ 

\hline
\hline

Mouse   &  3.1 $^{(a,b)}$ &  0.015 $^{(b)}$ &  880.0 $^{(a,b)}$  &  116.0 $^{(k)}$ 
&  7.59   \\
Rat     &  4.1 $^{(c)}$  &   0.014 $^{(i)}$ &  806.0 $^{(i)}$  &   71.4 $^{(l)}$  
&  11.29  \\     
Cat     &  5.1 $^{(d)}$  &   0.021 $^{(d)}$  &   780.0 $^{(j)}$  &   50.0 $^{(m)}$ 
&  15.60  \\
Dog     &  4.5 $^{(e)}$  &   ---     &    ---            &   12.6 $^{(m)}$  
&  ---    \\
Rhesus monkey   & 5.6 $^{(f)}$  &  0.009 $^{(f)}$ &  329.9 $^{(f)}$ &  37.4 $^{(n)}$ 
&  8.82  \\
Human   &  6.4 $^{(g,h)}$  &   0.023 $^{(g)}$  &   219.0 $^{(g)}$  &   38.5 $^{(n)}$  
&  5.69  \\

\hline

\hline
\end{tabular}
\end{center}
The ratio $\rho_{c}/\rho_{n}$ is the average capillary length per neuron. \\
References:
$^{(a)}$ \cite{boero};
$^{(b)}$ \cite{tsai};
$^{(c)}$ \cite{michaloudi};
$^{(d)}$ \cite{pawlik};
$^{(e)}$ \cite{luciano};
$^{(f)}$ \cite{weber};
$^{(g)}$ \cite{meier};
$^{(h)}$ \cite{lauwers};
$^{(i)}$ \cite{bar};
$^{(j)}$ \cite{tieman};
$^{(k)}$ \cite{braitenberg};
$^{(l)}$ \cite{houzel2006};
$^{(m)}$ \cite{haug};
$^{(n)}$ \cite{houzel2010}.

%\end{table}

\newpage

%\begin{table}
\begin{center}
%\caption{
{\bf Table S3:} Arterial partial oxygen pressure and average cortical CBF.
\begin{tabular}{| l l l l|}
\hline
\hline
Species &  arterial $p_{O2}$ [mm Hg] & CBF$_{cortex}$ 
&  CBF$_{cortex}$/$\rho_{n}$  \\

\hline
\hline

Mouse  &  93.0 $^{(a)}$  &  157.4 $^{(a)}$  &  1.36  \\ 
Rat    &  93.0 $^{(b)}$  &  131.8 $^{(b)}$  &  1.84  \\
Rabbit  &  75.0 $^{(c)}$   &  70.2 $^{(c)}$  &  ---   \\
Cynomolgus     &     &           &            \\
monkey   &  102.4 $^{(d)}$ &  57.9 $^{(d)}$  &  1.51  \\
Dog      &  107.0 $^{(e)}$ &  ---    &  ---   \\
Rhesus monkey   &  92.4 $^{(f)}$ &  52.5 $^{(f)}$ & 1.40  \\ 
Pig      &  104.0  $^{(g)}$ &  59.5 $^{(g)}$  &  ---  \\
Human    &  94.0  $^{(h,i)}$ &  42.6 $^{(j)}$  &  1.11 \\

\hline

\hline
\end{tabular}
\end{center} 
The ratio CBF$_{cortex}$/$\rho_{n}$ is the average cortical blood flow per neuron
($10^{-8}$ mL/min). \\
References:
$^{(a)}$  \cite{frietsch2007};
$^{(b)}$  \cite{frietsch2000};
$^{(c)}$  \cite{tuor};
$^{(d)}$  \cite{orlandi};
$^{(e)}$  \cite{heistad};  
$^{(f)}$  \cite{noda};
$^{(g)}$  \cite{delp}; 
$^{(h)}$  \cite{ito};
$^{(i)}$  \cite{hatazawa};
$^{(j)}$  \cite{bentourkia}.

%\end{table}

\newpage

\noindent{\Large \bf Appendix S1}

In this Appendix we want to show a formal step connecting Eq. (6) and Eq. (8)
in the main text. We want to find the exponent $\gamma$, which satisfies 
the following equation:

$\left(\ln(R_{c}/R_{o})\right)^{2/3} = (R_{c}/R_{o})^{\gamma}$. 

Taking the natural logarithm of both sides leads to 
$(2/3)\ln(\ln(R_{c}/R_{o})) = \gamma\ln(R_{c}/R_{o})$, which yields
Eq. (7) in the main text.

%\end{narrowtext}
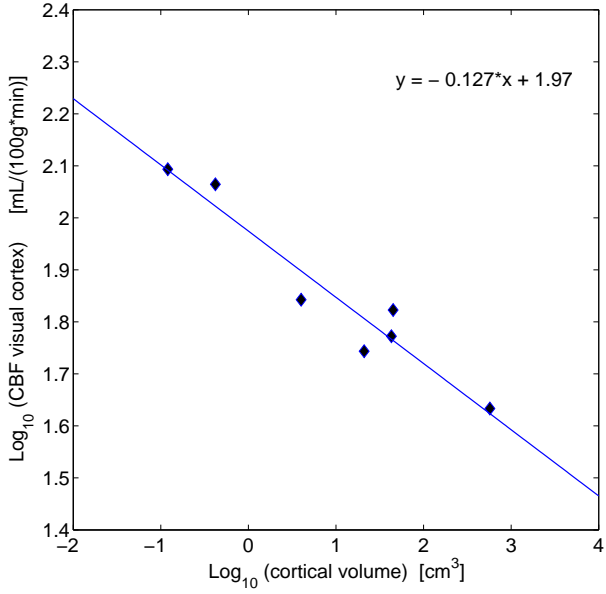
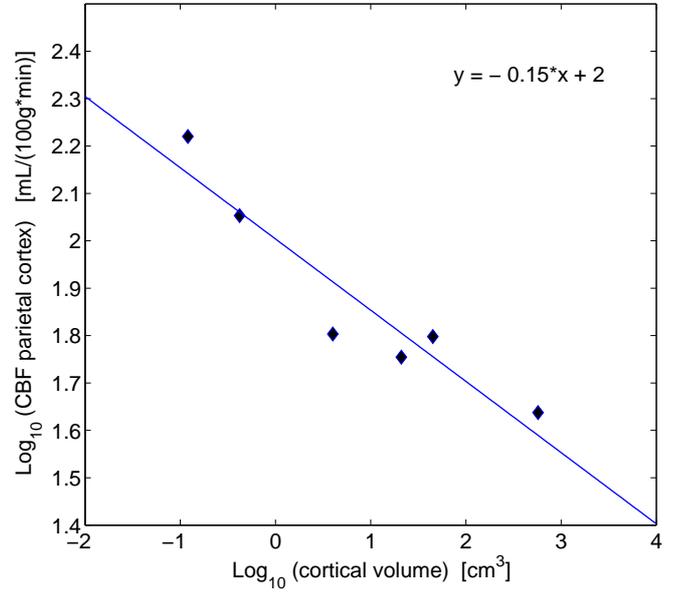
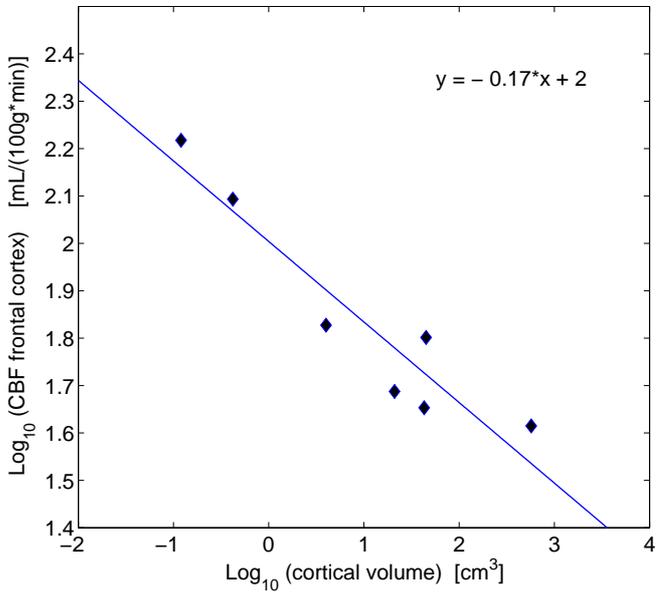
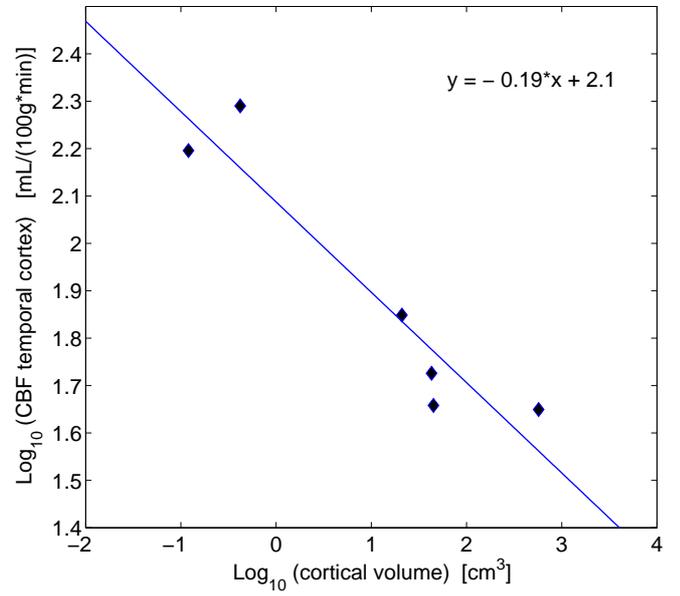
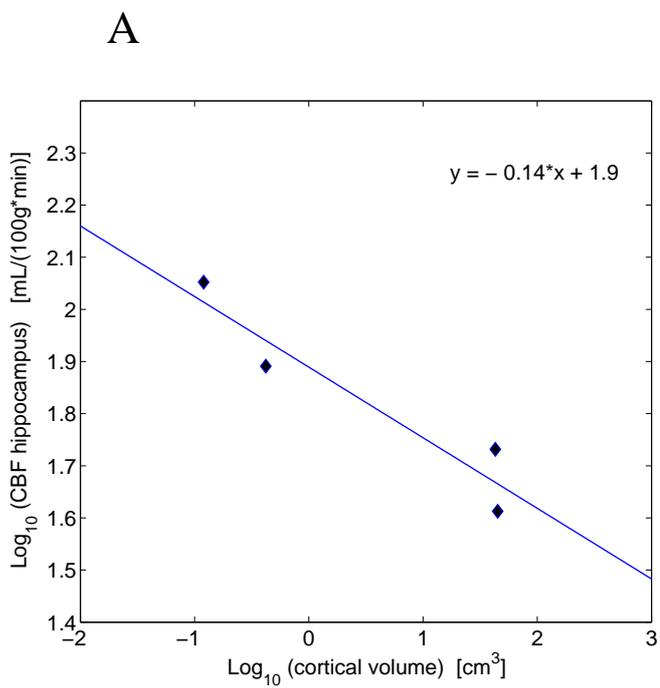
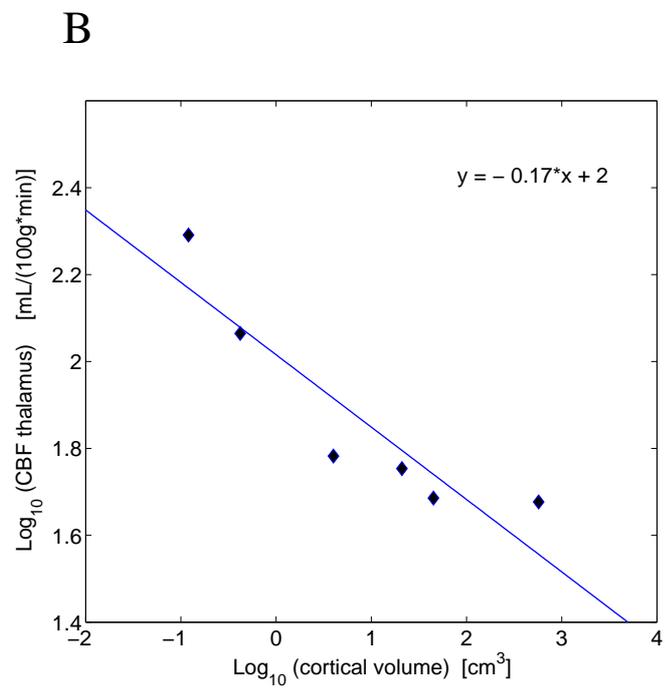
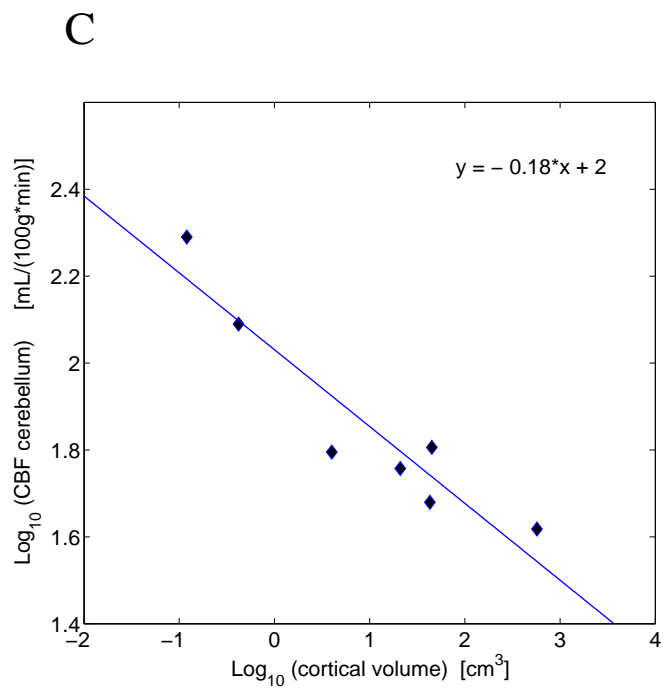
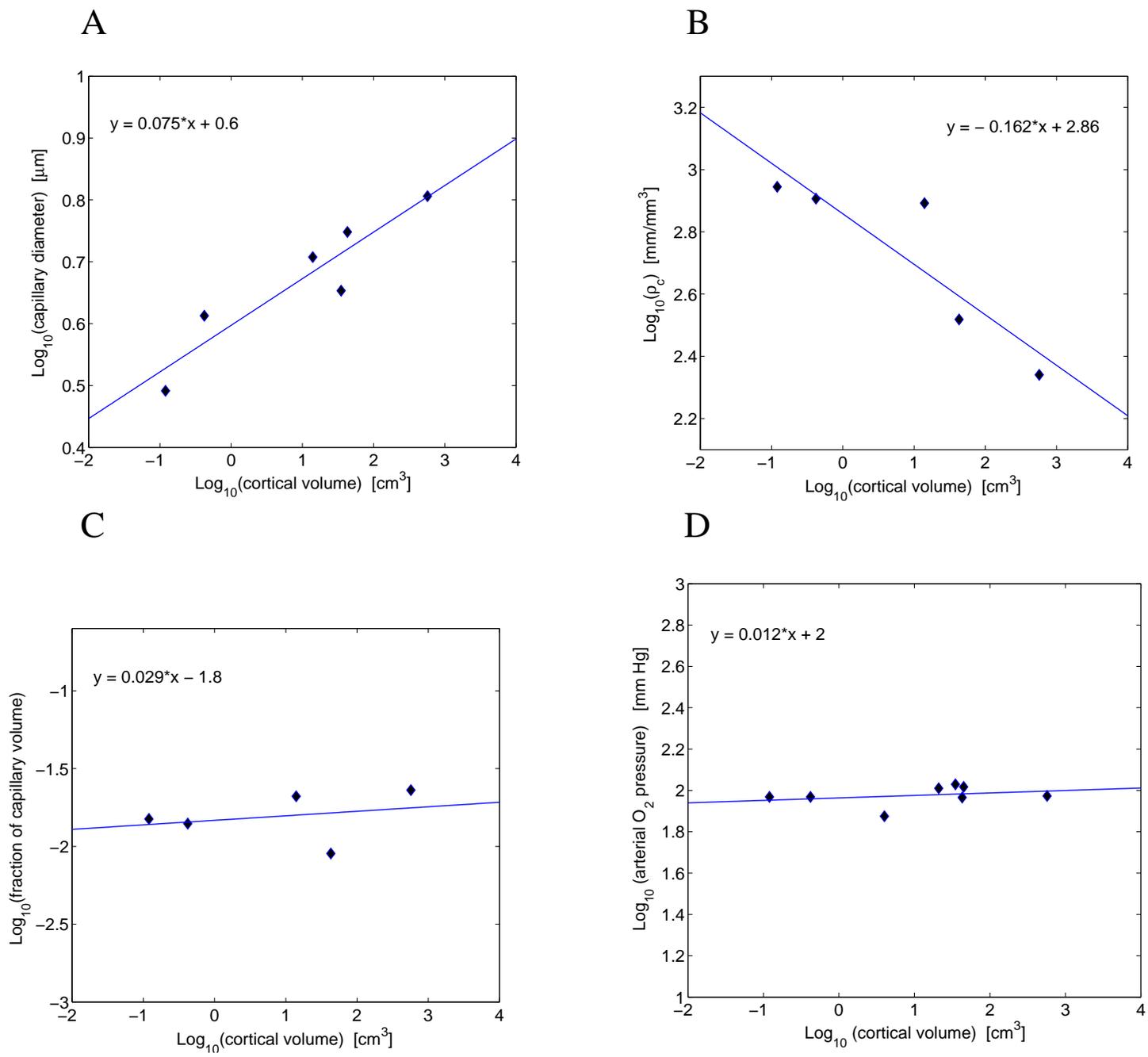
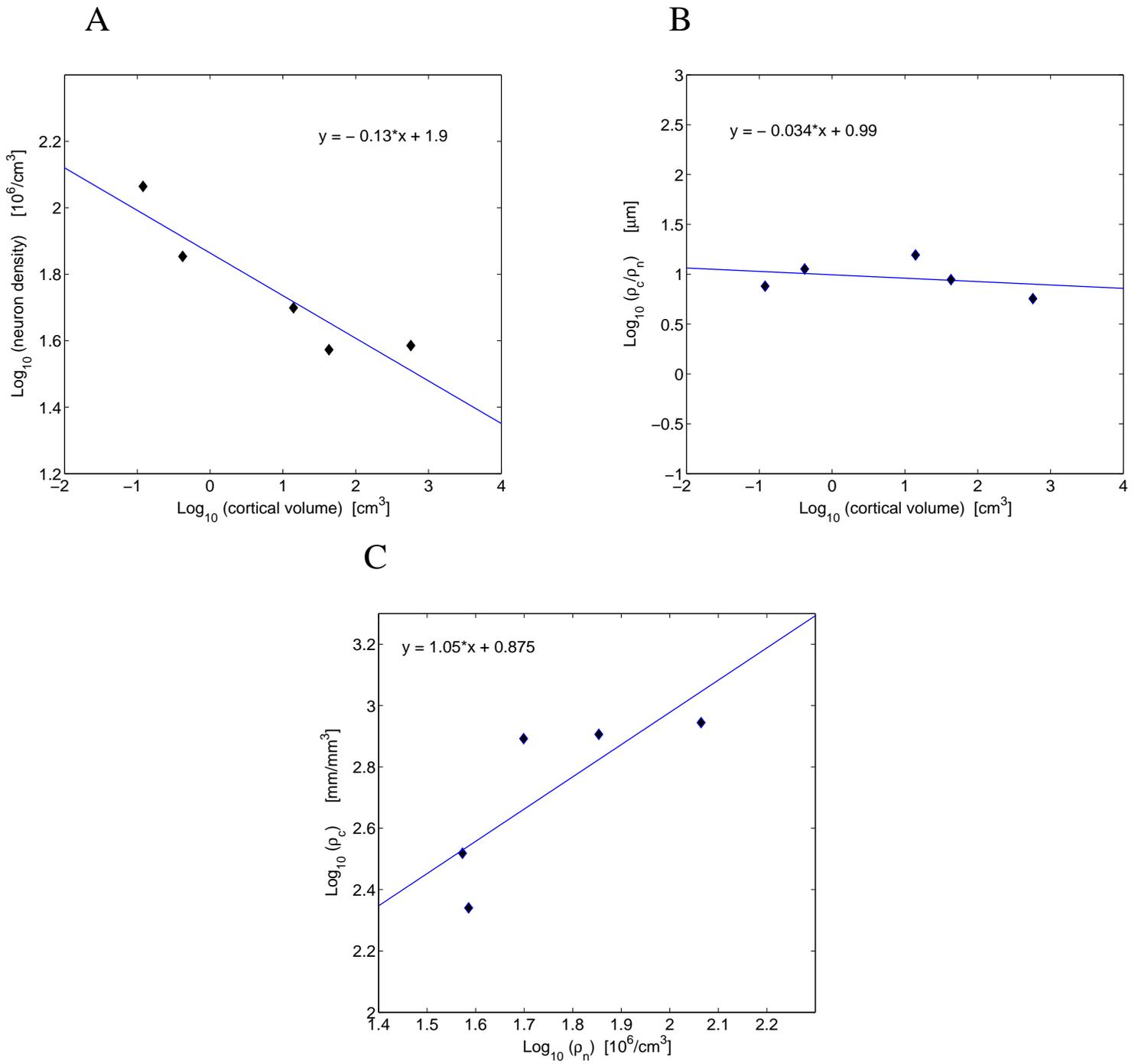
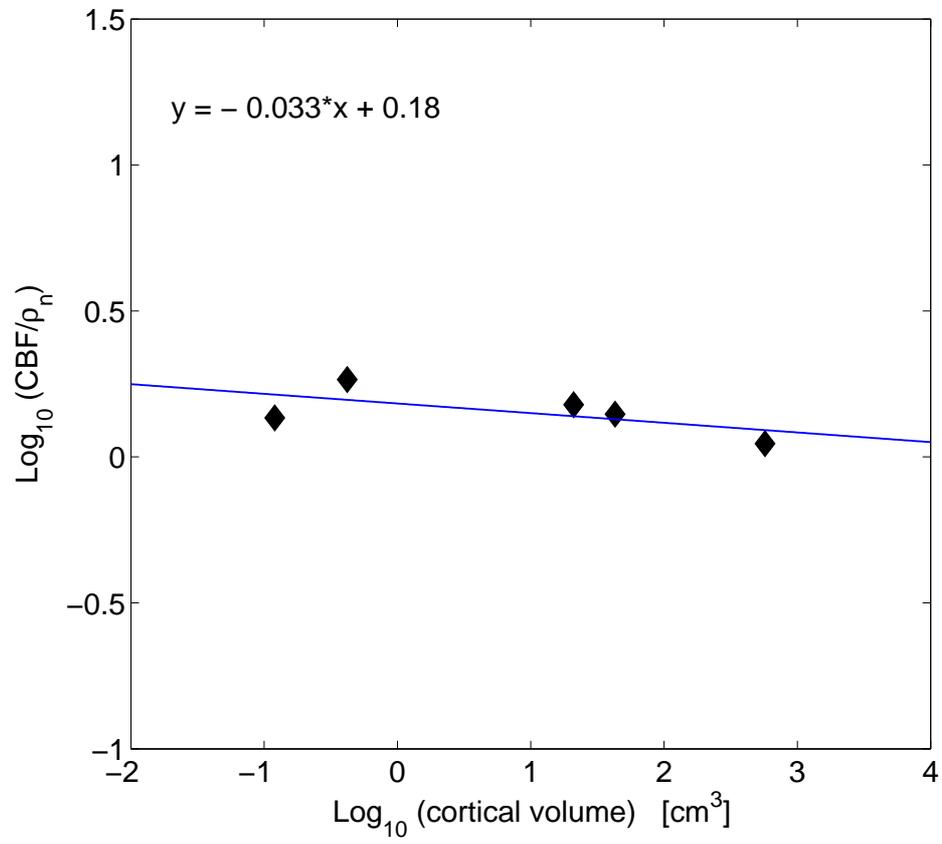
\end{document}